\documentclass[sigconf]{acmart}
\AtBeginDocument{%
  \providecommand\BibTeX{{%
    \normalfont B\kern-0.5em{\scshape i\kern-0.25em b}\kern-0.8em\TeX}}}

\setcopyright{acmcopyright}
\copyrightyear{2026}
\acmYear{2026}
\acmDOI{XXXXXXX.XXXXXXX}

\acmConference[IDE '26]{3rd International Workshop on Integrated Development Environments}{April 18,
  2026}{Rio de Janeiro, Brazil}

\acmBooktitle{IDE '26: 3rd International Workshop on Integrated Development Environments,
 April 18, 2026, Rio de Janeiro, Brazil} 
\acmISBN{978-1-4503-XXXX-X/18/06}

\def\Inst#1{\textsuperscript{#1}}
\def\Br{\\}
\makeatletter
\let\@mkbibcitationOrig\@mkbibcitation
\def\@mkbibcitation{\gdef\Br{ }\gdef\Inst##1{}\@mkbibcitationOrig}
\makeatother

\begin{document}

\title[Forecasting Developer Environments with GenAI: A Research Perspective]{Forecasting Developer Environments with GenAI:\\ A Research Perspective}

\author{Raula Gaikovina Kula}
\affiliation{
  \institution{The University of Osaka}
  \city{Osaka}
  \country{Japan}
}
\email{raula-k@ist.osaka-u.ac.jp}

\author{Christoph Treude}
\affiliation{
  \institution{Singapore Management University}
  \city{Singapore}
  \country{Singapore}
}
\email{ctreude@smu.edu.sg}

\author{Xing Hu}
\affiliation{
  \institution{Zhejiang University}
  \city{Hangzhou}
  \country{China}
}
\email{xinghu@zju.edu.cn}

\author{Sebastian Baltes}
\affiliation{
  \institution{Heidelberg University}
  \city{Heidelberg}
  \country{Germany}
}
\email{sebastian.baltes@uni-heidelberg.de}

\author{Earl T. Barr}
\affiliation{
  \institution{University College London}
  \city{London}
  \country{UK}
}
\email{e.barr@ucl.ac.uk}

\author{Kelly Blincoe}
\affiliation{
  \institution{University of Auckland}
  \city{Auckland}
  \country{New Zealand}
}
\email{k.blincoe@auckland.ac.nz}

\author{Fabio Calefato}
\affiliation{
  \institution{University of Bari}
  \city{Bari}
  \country{Italy}
}
\email{fabio.calefato@uniba.it}

\author{Junjie Chen}
\affiliation{
  \institution{Tianjin University}
  \city{Tianjin}
  \country{China}
}
\email{junjiechen@tju.edu.cn}

\author{Marc Cheong}
\affiliation{
  \institution{University of Melbourne}
  \city{Melbourne}
  \country{Australia}
}
\email{marc.cheong@unimelb.edu.au}

\author{Youmei Fan}
\affiliation{
  \institution{Nara Institute of Science and Technology}
  \city{Nara}
  \country{Japan}
}
\email{fan.youmei.fs2@is.naist.jp}

\author{Daniel M. German}
\affiliation{
  \institution{University of Victoria}
  \city{Victoria, BC}
  \country{Canada}
}
\email{dmg@uvic.ca}

\author{Marco Gerosa}
\affiliation{
  \institution{Northern Arizona University}
  \city{Flagstaff, AZ}
  \country{USA}
}
\email{Marco.Gerosa@nau.edu}

\author{Jin L.C. Guo}
\affiliation{
  \institution{McGill University}
  \city{Montreal, QC}
  \country{Canada}
}
\email{jguo@cs.mcgill.ca}

\author{Shinpei Hayashi}
\affiliation{
  \institution{Institute of Science Tokyo}
  \city{Tokyo}
  \country{Japan}
}
\email{hayashi@comp.isct.ac.jp}

\author{Robert Hirschfeld}
\affiliation{
  \institution{Hasso Plattner Institute \& University of Potsdam}
  \city{Potsdam}
  \country{Germany}
}
\email{robert.hirschfeld@hpi.uni-potsdam.de}

\author{Reid Holmes}
\affiliation{
  \institution{University of British Columbia}
  \city{Vancouver, BC}
  \country{Canada}
}
\email{rtholmes@cs.ubc.ca}

\author{Yintong Huo}
\affiliation{
  \institution{Singapore Management University}
  \city{Singapore}
  \country{Singapore}
}
\email{ythuo@smu.edu.sg}

\author{Takashi Kobayashi}
\affiliation{
  \institution{Institute of Science Tokyo}
  \city{Tokyo}
  \country{Japan}
}
\email{tkobaya@comp.isct.ac.jp}

\author{Michele Lanza}
\affiliation{
  \institution{Software Institute, USI, Lugano}
  \city{Lugano}
  \country{Switzerland}
}
\email{michele.lanza@usi.ch}

\author{Zhongxin Liu}
\affiliation{
  \institution{Zhejiang University}
  \city{Hangzhou}
  \country{China}
}
\email{liu\_zx@zju.edu.cn}

\author{Olivier Nourry}
\affiliation{
  \institution{The University of Osaka}
  \city{Osaka}
  \country{Japan}
}
\email{nourry@ist.osaka-u.ac.jp}

\author{Nicole Novielli}
\affiliation{
  \institution{University of Bari}
  \city{Bari}
  \country{Italy}
}
\email{nicole.novielli@uniba.it}

\author{Denys Poshyvanyk}
\affiliation{
  \institution{William \& Mary}
  \city{Williamsburg, VA}
  \country{USA}
}
\email{dposhyvanyk@gmail.com}

\author{Shinobu Saito}
\affiliation{
  \institution{NTT Computer and Data Science Laboratories}
  \city{Tokyo}
  \country{Japan}
}
\email{shinobu.saito@ntt.com}

\author{Kazumasa Shimari}
\affiliation{
  \institution{Nara Institute of Science and Technology}
  \city{Nara}
  \country{Japan}
}
\email{k.shimari@is.naist.jp}

\author{Igor Steinmacher}
\affiliation{
  \institution{Northern Arizona University}
  \city{Flagstaff, AZ}
  \country{USA}
}
\email{igor.steinmacher@nau.edu}

\author{Mairieli Wessel}
\affiliation{
  \institution{Radboud University}
  \city{Nijmegen}
  \country{Netherlands}
}
\email{mairieli.wessel@ru.nl}

\author{Markus Wagner}
\affiliation{
  \institution{Monash University}
  \city{Melbourne}
  \country{Australia}
}
\email{markus.wagner@monash.edu}

\author{Annie Vella}
\affiliation{
  \institution{University of Auckland}
  \city{Auckland}
  \country{New Zealand}
}
\email{annie.luxton@gmail.com}

\author{Laurie Williams}
\affiliation{
  \institution{North Carolina State University}
  \city{Raleigh, NC}
  \country{USA}
}
\email{lawilli3@ncsu.edu}

\author{Xin Xia}
\affiliation{
  \institution{Zhejiang University}
  \city{Hangzhou}
  \country{China}
}
\email{xin.xia@zju.edu.cn}

\renewcommand{\shortauthors}{Kula, Treude, Hu, et al.}

\begin{abstract}
Generative Artificial Intelligence (GenAI) models are achieving remarkable performance in various tasks, including code generation, testing, code review, and program repair.
The ability to increase the level of abstraction away from writing code has the potential to change the Human-AI interaction within the integrated development environment (IDE).
To explore the impact of GenAI on IDEs, 33 experts from the Software Engineering, Artificial Intelligence, and Human-Computer Interaction domains gathered to discuss challenges and opportunities at Shonan Meeting 222, a four-day intensive research meeting.
Four themes emerged as areas of interest for researchers and practitioners.
\end{abstract}




\maketitle

\section{Introduction}
Generative Artificial Intelligence (GenAI) models are already achieving remarkable performance in various software engineering tasks such as code generation, testing, code review, and program repair \cite{Deng2024,Xia2023,Yuan2024,gao2025current,pearce2025asleep}.
Furthermore, the rapid adoption of GenAI has compelled both practitioners and researchers to integrate these tools, often without fully understanding their long-term consequences~\cite{vaithilingam2022expectation,barke2023grounded}.

To understand the impact of GenAI on the Integrated Development Environment (IDE), 33 experts from the three domains of Software Engineering (SE), Artificial Intelligence (AI), and Human-Computer Interaction (HCI) gathered to discuss the grand challenges and opportunities at Shonan Meeting 222 ``\textit{The Future of Development Environments with AI Foundation Models}''\footnote{\url{https://shonan.nii.ac.jp/seminars/222/}}.
To help organize the Shonan, we sent out a pre-event survey that highlighted five questions that researchers felt were important when discussing the future of development environments:
\begin{itemize}
\item Which software development tasks should GenAI handle?
\item How should GenAI be integrated into IDE features?
\item What role remains for humans in software development?
\item Do we still need development environments?
\item Your boldest claim for software engineering in 2050?
\end{itemize}

Based on these five questions, participants then developed and presented their own impressions of what the future of IDEs would look like. 
As the presentations progressed, it was apparent that four themes emerged from these intensive discussions, which we discuss in the subsequent sections. 
It is important to note that participants did not all agree with the description of the themes, with several disagreements and divergent worldviews between the themes. For example, the participants were split between themes one and two. 

\section{Theme One: A Changing IDE, Solving the Same Problems}

Some researchers view AI-augmented IDEs as an evolutionary step rather than a radical overhaul of software engineering practice~\cite{ozkaya2023application,ozkaya2023next}. The core tasks—design, coding, debugging, testing—remain unchanged; GenAI simply automates the repetitive or mechanical aspects, freeing developers to devote more time to high-level design and creative problem-solving. This automation is expected to streamline workflows, reduce boilerplate work, and accelerate delivery cycles, thereby enhancing overall productivity~\cite{coutinho2024role,yu2025paradigm}.

Despite these benefits, significant hurdles persist. Aligning training data across the software engineering and AI domains demands careful curation, while divergent terminologies—such as symbolic versus neural representations~\cite{zhang2024bridging}—require interdisciplinary dialogue to create shared ontologies. Moreover, GenAI is increasingly taking care of low-level technicalities and idiosyncrasies, pushing abstraction layers higher within IDEs. This shift promises greater developer autonomy but also risks obscuring underlying implementation details if not managed thoughtfully~\cite{sergeyuk2025human,chen2025code,ulfsnes2024transforming,treude2025interactingaireasoningmodels}.

Ultimately, programming must remain an immersive, collaborative human activity. As GenAI handles more routine tasks, developers will be left to intervene only during rare yet critical moments—debugging complex issues or architecting scalable systems~\cite{houck2025space,kumar2025developers,pan2024don}, as recent research on developer-AI collaboration shows how interaction patterns are shifting~\cite{treude2025developers, settewong2025humandocumentaicode}. Ensuring that automation enhances rather than diminishes creativity requires IDE designs that encourage intentional interaction and maintain the developer's central role in decision-making \cite{Inman2025}.

\paragraph{\textbf{Lessons Learnt}}
The key lesson is that although the technologies are changing, the basic components of software development and its key guiding principles remain the same. 
We are simply moving to a higher level of abstraction, and all tedious work will be handled by the computing power, leaving tasks of critical thinking. 

\section{Theme Two: A New Paradigm That Begins from the IDE}

The integration of Generative AI into IDEs is expected to spawn a host of novel research avenues, spanning causal inference, legal implications, and software engineering education~\cite{nguyen2025generative}. By treating code as an artifact that can be questioned, challenged, and re-generated, these systems will enable developers to explore alternative implementations faster, fostering a turn-taking dynamic between human intent and AI suggestion~\cite{ferdowsi2024validating,Rojpaisarnkit2025}. This raises the need for specialized tools that support GenAI-driven code: knowledge and context management, orchestrated agent swarms executing complex workflows, automated checkpointing and repair mechanisms, and management of ``LLM technical debt'' that arises when large language models become tightly coupled to production systems~\cite{menshawy2024navigating,aljohani2025promptdebt}.

Moreover, as language models begin to generate increasingly sophisticated and self-correcting code, developers may find themselves acting more like digital gardeners than traditional programmers~\cite{qiu2025today,kula2025shiftwritingpruningsoftware}. Systems could evolve autonomously, with new tools emerging to address the unique problems posed by such self-evolving software—ranging from continuous verification of causality to ensuring compliance with evolving legal standards~\cite{weyns2023vision,cai2025ai}. In short, GenAI will not only augment existing IDE capabilities but also necessitate entirely new toolchains designed for a future where code is both created and maintained autonomously by intelligent agents with access to the right knowledge, ultimately realising a vision in which the IDE becomes genuinely integrated rather than the largely isolated environment it has often felt like so far~\cite{marron2024new,tufano2024autodev}.

\paragraph{\textbf{Lessons Learnt}}
Rather than just being an evolution, the IDE will cause us to rethink how we build software.
Ideas like evolving software might have to be replaced by on-demand software that can be easily thrown away.
The IDE will cater for a broader demographic of users, and software could be self-evolving without humans constantly intervening. 
On the other hand, this changing IDE might also bring new problems that were not previously faced by software developers, opening up new avenues for research.

\section{Theme Three: The Human Role Reimagined}

The contemporary developer increasingly assumes a managerial role in the software development ecosystem~\cite{kalliamvakou2017makes}. Rather than merely implementing code, developers must now grapple with \emph{comprehension debt}—the cognitive cost of understanding legacy systems, unfamiliar frameworks, and evolving requirements~\cite{agrawal2024code}. A personalized command center that aggregates relevant metrics, documentation, and contextual insights can mitigate this burden, allowing the developer to act as a project manager who orchestrates rather than writes code~\cite{marron2024new,hassan2024towards}.

In this paradigm shift, the human actor functions more as an \emph{author of intent}, a guide for the system's autonomous components, and a facilitator of collaboration among distributed agents~\cite{treude2025generative,rasheed2024autonomous,takerngsaksiri2025human}. The IDE should therefore support clarifying ambiguous specifications, enforcing quality assurance, and validating design decisions without requiring the developer to engage in low-level implementation details. By abstracting routine tasks, the environment frees developers to focus on higher-order problem-solving~\cite{marron2024new,hassan2025agentic}.

Design principles for such an IDE must prioritize assistance over frustration. Cognitive studies suggest that excessive mental load can impair creativity and lead to burnout~\cite{gonccales2019measuring,chandrasekaran2024enhancing}; consequently, interfaces should be intuitive, provide adaptive scaffolding, and offer clear pathways for intellectual exploration. Transparency and trust are essential: developers need to understand how the system arrives at its recommendations or predictions, which in turn fosters confidence in automated decisions~\cite{wang2024investigating,baltes2025rethinking}.

Ethical considerations also surface when humans delegate increasingly complex tasks to AI agents~\cite{dodig2025delegating,abrahao2025software}. Biases embedded in training data can propagate through decision-making pipelines, potentially harming users or reinforcing unfair practices~\cite{brun2018software,treude2023she,sami2023case}. The IDE should expose these biases and provide mechanisms for human oversight~\cite{atemkeng2024ethics}. Simultaneously, it must support continuous skills development, enabling developers to evolve from mere coders into system architects who design the overall structure rather than write every line of code~\cite{hassan2024towards,birillo2024bridging,sergeyuk2024design}.

Finally, a mutual \emph{theory of mind} between developer and AI is crucial: each party should model the other's intentions, constraints, and preferences~\cite{zhang2024mutual,wang2021towards}. This reciprocal understanding allows for more natural collaboration and reduces miscommunication. As the profession matures, the role of software engineers—sometimes dubbed ``source argonauts''—will shift: while their individual share of code may shrink, their influence over architectural decisions and strategic direction will grow, reflecting a new hierarchy within the development ecosystem~\cite{betz2025great,meade2019changing}.

\paragraph{\textbf{Lessons Learnt}}
How developers interact with the machine to build software will change. 
The relationship may evolve from simply giving instructions to now managing what the machine recommends.
This changing relationship may alter cognitive load and decision-making.
Thus the IDE will need to be robust to accommodate clear communication and verification of intentions, constraints, and preferences. 
Since building software is a collaborative activity, the IDE will also have to facilitate interactions between all members of the team, with some of them potentially not human.

\section{Theme Four: Futurecasting the 2050 IDE}

This \textit{futuring} exercise imagines the context of programming in the future: what can be hypothetically \emph{seen} in 2050, without being overly concerned with ``how to get there''.  

\paragraph{Immersion and User experience} Walking into the room, the system powers on and immediately fills the space. It clearly distinguishes itself from its surroundings yet remains immersive enough for a user or creator to recognize familiar functions; this suggests that developers of that era had reached some consensus about how invasive the environment should be to meet users' needs.
By waving their hands around, an entire simulation materializes, and intuitive controls allow navigation through the complex system. A time-travel feature lets a user simulate changes to understand their effects, while real-time visualization of collaborators shows how all these modifications are being introduced into the system.

\paragraph{Components of the IDE}
Digging into the artifacts of 2050, we find that they are equipped with multiple AI agents and multimodal interfaces. 
Among these is a super-AI that serves as the core CPU, orchestrating the execution of the artifact. By integrating these AIs, the environment can evolve and repair bugs autonomously without explicit lines of code from humans. The compiler is designed to detect ambiguities in natural language and to be syntax-agnostic for AI agents, ensuring that they run without errors.
The distinction between creating and using software no longer exists. People generate and modify software components directly through natural language, gestures, or biological and contextual signals. Every interaction with the system can alter its behavior, so users effectively become developers. The development environment interprets intent across multiple forms of input, allowing software to adapt instantly to new needs. Programming becomes a shared process between human and system, unfolding continuously through everyday use rather than within discrete development phases.

Taking inspiration from sci-fi franchises such as \emph{Star Trek} and \emph{Warhammer 40k}, the holodeck metaphor applies to the IDE of the future: ``wireless communion with the machine spirit''. What-if scenarios—e.g., ``What if Google Maps routed everyone down the same road?''—are answered using full-scale digital twins. In a boon for agility and rapid prototyping, dials for \emph{fidelity} and \emph{scope} allow rapid re-evaluation of the current scenario. Edge cases, exceptions, and unintended changes over time are best represented as clusters of unique outcomes; what is ``unintended'' will be revealed upfront if the system records everything. 

\paragraph{Feedback Loops in the IDE} Motivated by advances in quantum computing, many states of the system can be explored in parallel in a world where compute power is no longer a constraint, enabling abstractions and jumps when needed. Multiple users (or developers) can engage with the system together; tools exist for one participant to ``enter'' the system's internal state and make edits, thereby blurring the boundaries between creator and user.
From observing unintended outcomes in the simulator to steering the program toward more intended behaviors, an effective feedback loop is essential. As the system becomes increasingly complex, this loop must become diagnostic, helping developers reason about the immediate next step. Because the line between using and programming software blurs, the level of expertise required diversifies; consequently, the granularity and focus of the feedback are automatically adjusted to match each developer's technical background.

\paragraph{\textbf{Lessons Learnt}}
The key lesson of the exercise is that it might be time to let go of conventional notions of files and folder organization, version control, and source code as the primary artifact. A future beyond the keyboard as the main mode of input is possible, while collaboration between humans remains foundational to the IDE. 

\section{In Closing}
We forecast developer environments to undergo major changes, however, the themes reveal that our understanding of the direction of change is still far from set in stone. While prevailing trends suggest possible directions for their role, design and functionality, researchers must continually assess whether these trajectories advance scientific inquiry. This critical appraisal ensures that innovations align with the needs of both industry and academia rather than merely following market hype. 

There are other aspects of the IDE that were not discussed during the Shonan, but also deserve attention. 
For example, the legal aspects of using AI-driven IDEs generating all kinds of artifacts brings forth questions such as the responsibility if an AI-driven IDE creates erroneous code (or wrong designs or insufficient tests) that leads to severe consequences (e.g., loss of money, harm to humans). 
Still, we hope that these themes invite us to rethink what it means to ‘\textit{develop}’ software, with human creativity and judgment remaining central.

\bibliographystyle{ACM-Reference-Format}
\bibliography{ide26-shonan222}

\end{document}